\documentstyle[twocolumn,aps,prl,floats,psfig]{revtex}
\begin{document}
\draft 

\title{$\phi$-N Bound State}

\author{H. Gao,$^{1}$ T.-S.H. Lee,$^{2}$ V. Marinov$^{1}$} 
\address{
$^{1}$Laboratory for Nuclear Science and Department of Physics,\\ 
Massachusetts Institute of Technology, Cambridge, MA~02139, USA\\
$^{2}$Physics Division, Argonne National Laboratory, 
Argonne, IL 60439, USA}

\date{13 October 2000}

\maketitle

\begin{abstract}
We show that the QCD van der Waals attractive potential is strong 
enough to bind
a $\phi$ meson onto a nucleon inside a nucleus 
to form a bound state. The direct experimental 
 signature for such an exotic state is proposed in the case of 
subthreshold 
$\phi$ meson photoproduction from nuclear targets. 
The production rate is estimated and
such an experiment is found to be feasible at the Jefferson
Laboratory.
\end{abstract}


It has been suggested\cite{Brodsky:1990jd} that the QCD van der Waals
interaction, mediated by multi-gluon exchanges,
is dominant when the interacting two color singlet
hadrons have no common quarks. 
In fact, the QCD van
der Waals interaction is ehanced at low velocity as has been shown
by Luke, Manohar, and Savage\cite{Luke:1992tm}.
This finding supports the prediction that 
a nuclear-bound quarkonium can be produced in charm production 
reactions at threshold, and the interpretation that
the structures seen in
$s^{10} d\sigma/dt(p p \to p p)$ and the $A_{NN}$ spin correlation at
$\sqrt s \sim 5$ GeV and large cm angles \cite{Court:1986dh} can be 
attributed to $c \bar c uud uud$ resonant states\cite{Brodsky:1988xw}.
If these interpretations are correct, then
analogous effects could also be expected at the
strangeness threshold. The objective of this work is to explore
this possibility. 

We are motivated by the investigation of
the nuclear-bound quarkonium by Brodsky, Schmidt, and 
de T\'{e}ramond \cite{Brodsky:1990jd}. They used
a non-relativistic Yukawa type attractive
potential $V_{(Q\bar{Q})A}= -\alpha e^{-\mu r}/r$
characterizing the QCD van der Waals interaction. They determined the 
$\alpha$ and $\mu$ constants using the phenomenological model of
high-energy Pomeron interactions developed by Donnachie 
and Landshoff \cite{landshoff}.
Using a variational wave function 
$\Psi(r) = (\gamma^3 /\pi)^{1/2}e^{-\gamma r}$, they predicted bound
states of $\eta_{c}$ with $^3$He and heavier nuclei. Their prediction
was confirmed by Wasson \cite{wasson} using a more realistic 
$V_{(Q\bar{Q})A}$ potential taking into account the nucleon
distribution inside the nucleus.

Similarly, one expects the attractive QCD van
der Waals force dominates the $\phi$-N interaction 
since the $\phi$ meson is almost a pure $s\bar{s}$ state.
It is possible that a $\phi$-N bound state or resonant state can be
formed in some reactions.
In photoproduction of $\phi$ meson from a proton target above 
threshold, the formation of a bound $\phi$-N state is not likely
because of the momentum mismatch between the $\phi$ and the recoil
proton. As such, no experimental evidence exists on the formation of
the $\phi$-N bound state up to now. On the other hand,
such a $\phi$-N bound state could be formed 
inside a nucleus. In this paper, we will verify this possibility and
make predictions for future experimental tests.

Using the variational method and following Ref.\cite{Brodsky:1990jd} to
assume $V_{(s\bar{s}), N}=-\alpha e^{\mu r}/r$, we find that a bound state of 
$\phi$-N is possible with $\alpha=1.25$   
and $\mu= 0.6$ GeV. The binding energy obtained is 1.8 MeV. 
Our results should be compared with $\alpha=0.6$ and
$\mu=0.6$ GeV determined by
Brodsky, Schmidt, and 
de T\'{e}ramond \cite{Brodsky:1990jd} for the $c\bar{c}$ quarkonium.
The interaction is expected to be enhanced
by $(m_c/m_s)^3$, i.e., $q\bar q$ separation cubed, from $c\bar{c}$
to $s\bar{s}$.
Since the radius of the $\phi$ meson 
is (0.4 fm \cite{povh}) twice
the radius of the $J/\Psi$ meson, $\alpha = 1.25$ is a rather
conservative coupling
constant to use for the $\phi$-N interaction. 
Also, the interaction is expected to have longer range for 
the $\phi$-N system than that of the $c\bar{c}$-N interaction.
Thus, $\mu$ = 0.6 GeV used in the variational approach described above
is also conservative for the $\phi$-N interaction.   
Further, this attractive force is enhanced inside the nucleus in the 
quasifree subthreshold photoproduction kinematics.
Hence, a bound state of $\phi$-N is possible from the attractive 
QCD van der Waals force.
Our qualitative finding can be verified on the lattice in the 
future \cite{liu}.

Experimentally, it is possible to observe the formation of a bound state 
$\phi$-N in the subthreshold quasifree 
$\phi$ photoproduction process.
The incoming photon couples to a moving nucleon 
inside the nucleus.  
The $\phi$ meson can be produced near rest inside 
the nuclear medium in the laboratory frame when the initial nucleon is
moving in a direction opposite to that of the incoming photon.
The attractive QCD van der Waals force between the $\phi$ meson 
and nucleons
inside the residual nuclear system enhances the probability for 
the formation of the $\phi$-N
bound state.
 It is thus possible for a bound state
$\phi$-N to be formed inside the nucleus in the 
quasifree subthreshold $\phi$ production process from a 
nuclear target.
The experimental search for such a bound state would be a triple
coincidence detection of kinematically correlated $K^{+}$, $K^{-}$,
and proton in the final state.
The momentum
distributions of these final state particles are different from
those from unbound quasi-free $\phi$ production and the direct
quasifree $K^{+}K^{-}$ production.
Thus, it is possible
to identify a bound $\phi$-N state experimentally using the above
mentioned triple coincidence measurement.    
Such an experiment is feasible at Jefferson Laboratory
 where advantages  
of the continuous-wave electron beam, high luminosity, and the 
state-of-the-art detector package will be utilized to their full capabilities.
The rate estimate for such a measurement is described below.


We assume that the photoproduction of a $\phi$-N bound state,
called $d$, from nuclei is a two-step process
and can be evaluated in the impulse approximation. The reaction
mechanism is illustrated in Fig.~1. We consider the production on
a p-shell nucleus, like $^{12}C$, and assume that its structure
can be described by the simple shell model
with harmonic oscillator wavefunctions.
By using the closure to sum over the
intermediate $(A-1)$-nucleon states, the reaction amplitude for
$A_i(\gamma,d)A_f$ can be written in the rest frame
of the initial nucleus $A_i$ as
\begin{eqnarray}
T_{fi}(p_d,p,;q,E) &=& \int d\vec{k}
[\sum_{\alpha\neq \beta}\phi_\alpha(\vec{p}+\vec{k}-\vec{q})
\phi_\beta(\vec{p}_d-\vec{k})]
\nonumber \\
&\times& F(p_d;k,(p_d-k))\nonumber \\
&\times& \frac{1}{E-E_{A-1}(\vec{q}-\vec{p}-\vec{k})
-E_\phi (k)-E_N(p) + i\epsilon}\nonumber \\
&\times& t(k,p;q,(p+k-q))
\end{eqnarray}
where $E_\alpha(k)=[m_\alpha^2+\vec{k}^2]^{1/2}$ is
the energy for the particle $\alpha$ with momentum
$\vec{k}$, $t(k,p;q,p_1)$
 is the amplitude for the $\gamma(q) + N(p_1 ) \rightarrow
\phi(k) + N(p)$ transition, $F(p_d;k,p_2)$ is
for the $\phi(k) + N(p_2) \rightarrow d(p_d)$ 
transition, and $\phi_{\alpha}(p)$ is the normalized harmonic
oscillator wavefunction.
The total energy is $E=E_i+q$, and $E_i$  denotes the energy
of the initial nucleus. Noting that $p_1=p+k-q$ and $p_2 = p_d-k$,
Eq.(1) can be easily identified with Fig.1.

\begin{figure}[t]
\psfig{file=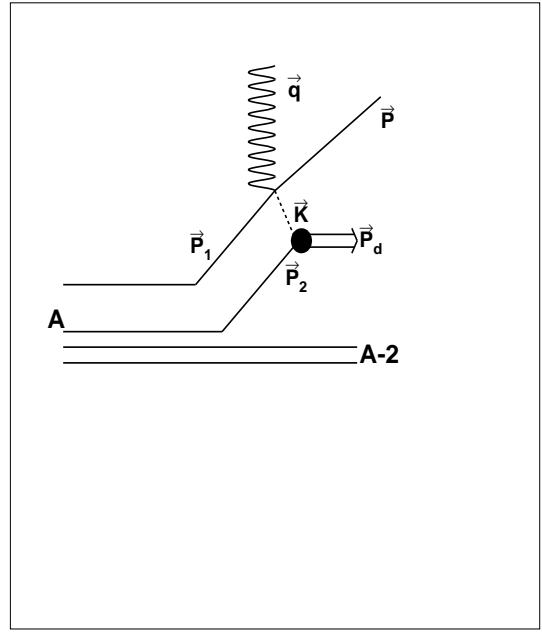,height=10.cm,width=8.cm}
\caption[]{The graphical representation of 
the quasifree process discussed in the text.} 
\label{diagram}
\end{figure}

We consider the energies near and below the $\phi$ production 
threshold. It is
then reasonable to assume 
that the intermediate $\phi$-N is in s-wave and
its transition matrix elements can be simplified as 
\begin{eqnarray}
t(k,p;q,p_1) &\sim& \frac{1}{4\pi}t_0(Q,q_c), 
\end{eqnarray}
where $t_0(Q,q_c)$ is the on-shell $\gamma N \rightarrow \phi$-N amplitude
in the center of mass frame of the 
$\gamma N$ subsystem, 
$q_c$ is the relative momenta of the $\gamma N$ 
subsystem, and $Q$ is the relative momentum of the $\phi$-N
subsystem. 
 We can estimate $t_0(Q,q_c)$ from the total
cross section of $\gamma N \rightarrow \phi$-N defined by 
\begin{eqnarray}
\sigma^{tot}(\omega)=\frac{4\pi}{q_c^2} \rho_{Q}
\mid t_0(Q,q_c)\mid^2 \rho_{q_c} 
\end{eqnarray}
where $\omega=q_c + E_N(q_c) = E_\phi(Q) + E_N(Q)$,
 the density of states are
$\rho_{q_c} = \pi q_c^2E_N(q_c)/(q_c+E_N(q_c))$ and
$ \rho_{Q}=\pi QE_\phi(Q)E_N(Q)/(E_N(Q)+E_\phi(Q))$.
We use the $\sigma^{tot}(\omega)$ predicted by the model of 
Ref.\cite{titovlee} which was constructed to fit the available
data. We find that the $\sigma^{tot}(\omega)$
from threshold up to about 1.7 GeV can be fitted by
$t_0(Q,q_c)\sim 0.5 \times 10^{-8}$ (MeV)$^{-2}$.

For a s-wave intermediate $\phi$-N state, the $\phi N \rightarrow d$
amplitude can be written as
\begin{eqnarray}
F(p_d;k,p_2)  = \frac{\sqrt{4\pi}}{(2\pi)^{3/2}}
\int r^2 dr \frac{sin(Q^\prime r)}{Q^\prime r} 
[ e^{-\Lambda^2 Q^{\prime 2}} V_{(s\bar{s}) N}(r)] \Psi(r)
\end{eqnarray}
where $Q^\prime$ is the relative momentum of the $\phi$-N subsystem,
  the van der Walls potential $V_{(s\bar{s}) N}(r)$ and the
normalized wavefunction $\Psi(r)$ for the bound state $d$ have been
determined from a variational calculation, as discussed above.
Here we have introduced a cutoff function $e^{-\Lambda Q^{'2}}$ to 
assure that the bound $\phi^0$-N system can be formed 
by the van der Walls potential mainly in the region that
the relative motion between $\phi^0$ and N is slow. 

Neglecting the recoil energy of the final nucleus, 
the differential cross section of $A_i(\gamma,d)A_f$  can be
calculated from the reaction amplitude defined above 
\begin{eqnarray}
\frac{d\sigma}{dp_d d\Omega_d} \sim
(2\pi)^4 p_d^2E_N(p)p\int d\Omega_p \mid T_{fi}(p_dp;q,E)\mid ^2
\end{eqnarray}
where $p$ is evaluated from energy conservation
$q+Am_N=E_N(p)+E_d(p_d)+(A-2)m_N$ and the binding energy is neglected.

We have applied the above formula to calculate the photoduction of
a $\phi$-N bound state $d$ from $^{12}C$. The oscillator wavefunction
parameter is chosen to be $b =1.64 $ Fermi.
Fig.~2 shows the calculated
total cross section as a function of the photon energy and the cutoff 
parameter, $\Lambda$. 
For a given $\Lambda$ value, the total cross section peaked around
a photon energy of 1450 MeV, below the free production threshold of
1570 MeV. This peaking of the cross section below the threshold is
expected because of the enhancement of the Van del Waals attractive
force when the relative velocity between the intermediate $\phi$
and nucleons in the residual nuclear system is smallest. 
The cross section drops as the photon energy increases and reach 0.008 nb
at $E_\gamma = 1800$ MeV. 
At a photon energy $E_\gamma = 1450 $ MeV, 
the calculated total cross
section is found to be $\sigma^{tot}= 1.4 $ nb for a rather large 
cutoff $\Lambda = 3$ Fermi.
Thus the production of the $\phi^0$-N is only
accessible at energies below and near threshold. 

\begin{figure}[t]
\psfig{file=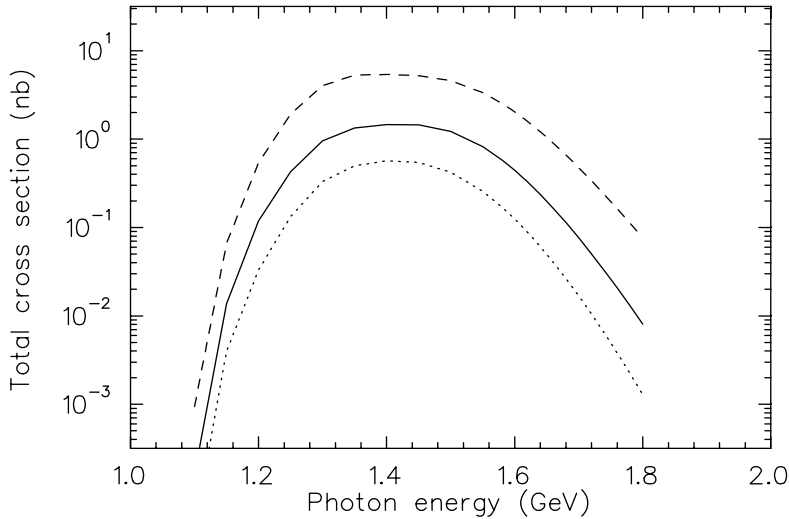,height=10.cm,width=7.cm}
\caption[]{The calculated total cross section for the formation of the
$\phi$-N bound state as a function of the photon energy. The
dashed, solid and the dotted curves are for $\Lambda$ = 2,3,4 Fermi 
correspondingly.} 
\label{phi_plot}
\end{figure}

Based on the total cross section calculated above, 
the search for the $\phi$-N bound state is feasible experimentally
by the triple coincidence detection of proton, $K^{+}$, and $K^{-}$ as
described previously.
Using a large acceptance detector system and a luminosity of
$10^{35}$/cm$^2$/sec from an untagged photon beam at Jefferson Laboratory, 
the triple coincidence event from the $\phi$-N bound state decay
is estimated to be about 180/hour at a subthreshold kinematics 
with an incident photon energy of 1450 MeV. 
An average flight path of 2 meters
for kaon detection, and realistic detector efficiencies were used
in the above rate estimation.

In the case of the $J/\psi$-nucleon scattering, 
Brodsky and Miller \cite{miller} conclude that the 
gluonic van der Waals interaction dominates the scattering. The hadronic 
corrections to gluon exchange which are generated by $\rho\pi$ and 
$D\bar{D}$ intermediate states of the $J/\psi$ are shown to be negligible.
For the $\phi$-nucleon system, the ``intrinsic strangeness'' in 
the nucleon may complicate the simple picture of gluonic van der Waals 
interaction being the dominating interaction because of the possible strange
quark exchange contribution. 
Thus,
the experimental search for the $\phi$-N bound state will 
not only help to unveil the nature of the QCD
van der Waals force, but may also help to probe the strangeness content of the
nucleon.

In conclusion, we found that the QCD Van der Waals attractive force is
strong enough to form a bound $\phi$-N state inside the nucleus. 
Experimentally, it is possible to search for such a bound state using
the $\phi$ meson below threshold quasifree photoproduction kinematics.
Using a simple model, we calculated the rate for such subthreshold quasifree
production process using a realistic Jefferson Laboratory luminosity and a large
acceptance detection system. We conclude 
that such an experiment is feasible.

We thank A. Bernstein, S.J. Brodsky, N. Isgur, K.-F. Liu, R.J. Holt,
B.-Q. Ma, and 
G.A. Miller for stimulating discussions.
We thank the hospitality of the National Taiwan University 
where part of this work was carried out.
This work was supported   
by the U.S. Department of Energy under 
contract number DE-FC02-94ER40818,  and also by
 U.S. Department of
Energy, Nuclear Physics Division, contract No. W-31-109-ENG-38.

\end{document}